\documentclass{epl}
\usepackage{epsfig,amssymb,amsfonts,amsmath}
\usepackage{graphicx}
\title{Phase coexistence and relaxation of the spherical
frustrated Blume-Emery-Griffiths model with attractive
particles coupling}
\shorttitle{Mean field spherical frustrated BEG}
\author{A. Caiazzo, A. Coniglio, M. Nicodemi}
\institute{Dipartimento di Scienze Fisiche, INFM, Unita' di Napoli,\\ 
Monte Sant'Angelo, I-80126 Napoli, Italy}
\newcommand{\dint}{\displaystyle \int}

\begin{document}

\maketitle

\begin{abstract}
We study the equilibrium and dynamical properties of a  spherical version of
the frustrated
Blume-Emery-Griffiths model at mean field level
for attractive particle-particle coupling ($K>0$).
Beyond a second order transition line from a paramagnetic to a 
(replica symmetric) spin glass phase, 
the density-temperature phase diagram is characterized by
a tricritical point from which, interestingly, a first order transition
line starts with coexistence of the two phases. In the Langevin dynamics
the paramagnetic/spin glass discontinuous transition line is found to be dependent on the
initial density; close to this line, on the paramagnetic side, the correlation-response
plot displays interrupted aging. 
\end{abstract}

Various mean field spin glass models have been introduced and investigated
in the last years;
recently, a few models have
been introduced \cite{GS,IFLG,BEG,CL} which combine features of spin glasses 
and lattice gas, and thus allow to describe important physical quantities, 
such as density.
The frustrated Blume-Emery-Griffiths (BEG) model \cite{BEG,CL},
that we discuss here, is a general framework of such kind.
Essentially 
it consists of a lattice gas ($n_i=0,1$) where 
the particles interact through their
internal degree of freedom ($s_i=\pm 1$) via quenched random couplings
and also through a potential depending on a nonrandom coupling constant, $K$. 
A complete mean field equilibrium solution of the frustrated BEG has been recently found
\cite{CL} by numerically solving the equations of the full replica
symmetry breaking scheme through a suitable technique of integration.
The solution has given evidence of a new interesting behavior
in the spin glass panorama, namely a 
coexistence line between a paramagnetic ($P$) and a
spin glass ($SG$) phase;
the dynamics along this line has not yet been investigated.
In this letter we study a spherical version of the
mean field frustrated BEG model in the attractive case ($K>0$) 
that can be analytically solved in its equilibrium and Langevin
dynamics.  

The equilibrium phase diagram we get essentially 
reproduces
that of the Ising version ($s_i=\pm 1,n_i=0,1$) of the model \cite{CL}:
the plane density-temperature shows
a second order transition line
from a $P$ to a (replica symmetric) $SG$ phase,
ending in a tricritical point;
here, a first order
transition line starts where coexistence of the two phases exists.
The tricritical point lies on the low density branch of the $P$ spinodal ($K<1$, Fig. 1a),
or on the side of the high density branch ($K>1$, Fig. 1b); in this last case both a $P/P$
and a $P/SG$ first order line are obtained.
In the Langevin dynamics from a random initial condition (rapid quench from high temperature)
the equilibrium correlation functions display a diverging relaxation time
at the second order line. 
Furthermore, the $P/SG$ (and, if present, the $P/P$) discontinuous transition
line is found to be dependent on the initial density $d\left( 0\right) $ (Fig. 3);
this critical line
does not correspond to any singularity of the equilibrium correlation
functions, but rather to a diverging plateau displayed by the density function
$d\left( t\right) $,
when its long time limit jumps from an asymptotic (equilibrium) value to the
other (Fig. 2).
One can recognize the glassy behavior from the two-time quantities
which exhibit an aging regime in the whole glassy phase; furthermore,
still in the nonglassy phase, close enough to the $P/SG$ 
discontinuous transition line, one can explicitly observe the pattern
of interrupted aging from the correlation-response diagram (Fig. 4).

\smallskip
Our spherical version of the frustrated mean field BEG model is defined as follows.
Let us consider the
Ising version of the model, described by the Hamiltonian
\begin{equation}
H=-\sum_{i<j}J_{ij}s_{i}n_{i}
s_{j}n_{j}-\frac KN\sum_{i<j}n_{i}
n_{j}-\mu \sum_i
n_{i}  \label{ham}
\end{equation}
$\mu $ being the chemical potential and $J_{ij}$ quenched Gaussian couplings with zero mean and variance $%
\left[ J_{ij}^2\right] =1/N$. 
We introduce two new spin fields $s_{1i},s_{2i} $ 
through the change of variables $s_i=s_{1i}$, $n_i=\left( s_{1i}s_{2i}+
1\right) /2$. 
Then, we assume
$s_{1i},s_{2i} $ as continuous variables, subjected to spherical constraints ($\sum_is_{ai}^2=N$ for $%
a=1,2$). Hereafter we shall use the variables $s_{ai}$ to describe the model.
The statics can be solved using the eigenvalue density $\rho
\left( \lambda \right) $ of a large random matrix
\cite{replica}. 
The $P$
and the $SG$ phase are described by:  
\begin{equation}
\begin{array}{llll}
\dfrac {z-\sqrt{%
z^2-4\beta ^2}}{2\beta ^2} =d\quad \quad &
d=1-\dfrac 1{z+2\beta \mu +2\beta Kd}\quad \quad & q=0 & P \\ 
z=2\beta &
d=1-\dfrac 1{2\beta +2\beta \mu +2\beta Kd} & q=d-\dfrac 1\beta \quad \quad & SG
\end{array}
\label{P-sd.eq2}
\end{equation}
where
$d=\left[
\left\langle s_{1i}s_{2i}\right\rangle +1\right] /2$ denotes the
density, 
$q=\left[ \left\langle
s_{1i}+s_{2i}\right\rangle ^2\right] /4$ the overlap and
$z=z_1-\beta \mu -\beta Kd$
with $z_1=z_2$ the two Lagrange
multipliers related to the spherical constraints.
The stability analysis
yields two critical lines,
called spinodal lines, at which 
the corresponding solution becomes unstable 
against density fluctuations
(and both the compressibility
and the specific heat
are found to diverge): 
\begin{equation}
T=\left( -K+\sqrt{K^2+\left( 1-d\right) ^{-2}+d^{-2}}\right) ^{-1}\,
\text{%
$P$-spinodal};\quad T=2K\left( 1-d\right) ^2\,\,
\text{$SG$-spinodal}
\label{sp.lines}
\end{equation}
A second order transition line
is located by $d=T$ in the plane $d-T$ (or by $T=T_c\left( \mu ,K\right) $
in the plane $\mu -T$); below this line 
one finds from the first of Eqs. (\ref{P-sd.eq2}) that the $P$ solution 
becomes unphysical. By increasing the parameter $K$ the spinodal lines move
to higher temperatures,
while the second order line is $K$-independent.  
For $K<1$ (Fig. 1a) the second order line meets the spinodal lines in 
a tricritical point $A$, with temperature $T_{A}$, located on the low density
branch of the $P$ spinodal: 
\begin{equation}
T_c\left( \mu
,K\right) =\frac{2K-3-2\mu +\sqrt{\left( 2K-3-2\mu \right) ^2+16K\left(
\mu +1\right) }}{4K};\quad
T_{A}\left( K\right) =\frac{4K+1-\sqrt{8K+1}}{4K}
\label{sec.ord.line}
\end{equation}
Below the point $A$, 
the system undergoes a $P/SG$ first order transition
where the order parameters $q$ and $d$ discontinuously jump and a
latent heat is involved; the critical
line is numerically determined
by imposing the free energy balance between the two
solutions. Between the spinodal lines in the plane $d-T$, no pure phase is
achievable, not even as a metastable one, and the system is a mixture of the
two phases. 
In the zero $K$ limit, $T_A$ vanishes 
and the first order line disappears \cite{CCN}. 
For $K>1$ (Fig. 1b) the second order line intersects the $P$ spinodal
in its high density branch; in this case a first order transition from 
the low to the high density $P$ phase is also exhibited.  
Analytical expressions of the thermodynamical quantities can be
derived in terms of the above solutions, Eqs. (\ref{P-sd.eq2}).
The linear susceptibility, the compressibility and the specific heat
 show a cusp by crossing the second order line \cite{CCN}
and a finite jump at
the first order ones.
The $SG$ susceptibility is infinite in the whole $SG$ phase, while remains finite
in the $P$ phase, diverging when
the second order line is approached \cite{CCN}.
\vspace{0.5cm}

\begin{figure}[h!]
\begin{center}
\vspace{-.6cm}
\hspace{-.4cm}
\epsfxsize=8cm
\epsfysize=7.5cm
\epsfbox{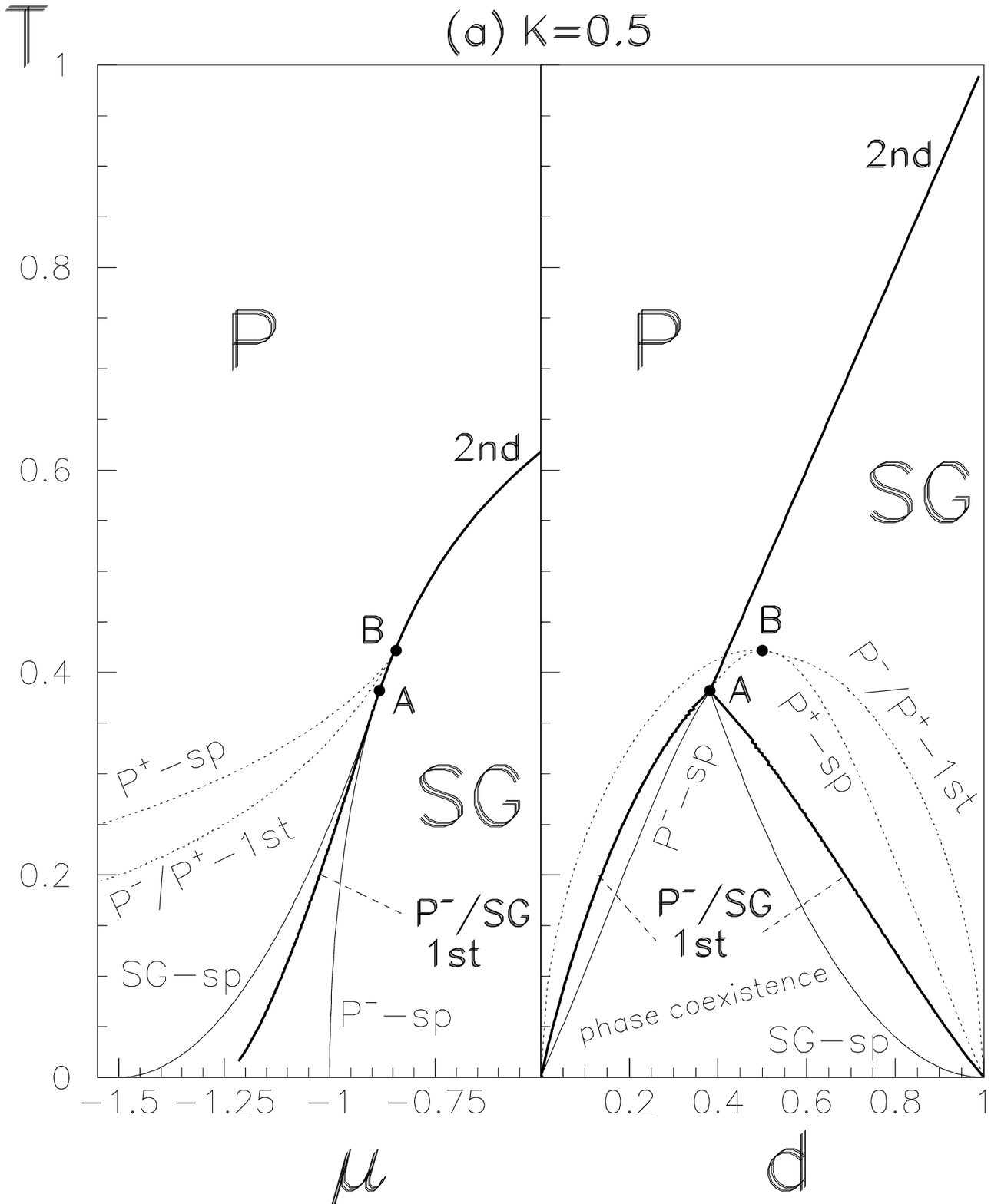}
\hspace{-1.7cm}
\epsfxsize=8cm
\epsfysize=7.5cm
\epsfbox{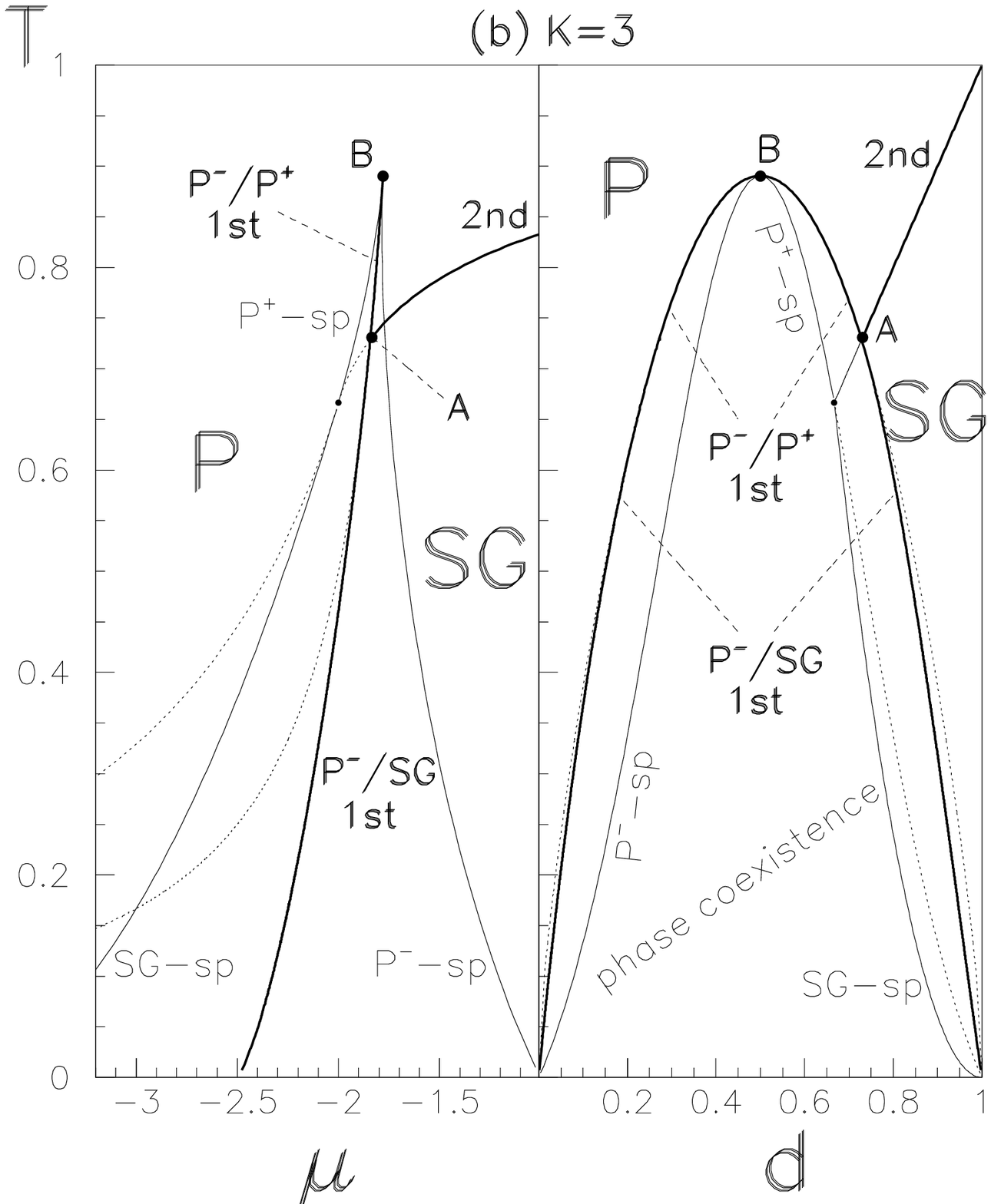}
\end{center}
\vspace{-0.5cm}
\caption{\footnotesize 
Equilibrium phase diagrams in the plane $\mu - T$ and $d-T$ for $K=0.5$ 
$\boldsymbol{(a)}$ and $K=3$ $\boldsymbol{(b)}$. One can have a
spin glass ($SG$) phase or a paramagnetic ($P$) phase (which forks at the point $B$ in
a low density $P^-$ and an high density $P^+$ component). 
The transition lines are marked in bold; the spinodal lines,
at which the relative phase becomes unstable, are also shown as continuous lines; dotted lines 
represent unphysical analytical continuations of critical lines.
Finally, full circles mark the tricritical point $A$ and the bifurcation point $B$ on the
$P$ spinodal line. Notice that the $P$ phase becomes unphysical below the 
second order line in the plane $d-T$; a $P/SG$ first order transition line 
starts out from the tricritical point $A$,
located on the $P^{-}$ spinodal in the case $\boldsymbol{(a)}$ or on the side of the 
$P^{+}$ spinodal
in the case $\boldsymbol{(b)}$.   } 
\label{beg3}
\end{figure}

Now we deal with the dynamics.
We
assume that the two spin fields $s_{ai}$ evolve via usual Langevin equations; using
the mean field approximation we can replace the
particle-particle interaction
with $-Kd\left( t\right) \sum_i\left( s_{1i}s_{2i}+1\right) /2$. Thus when projected
onto the basis $\lambda $ where $J_{ij}$ is diagonal, the Langevin equations
become: 
\begin{equation}
\frac {d}{dt}\left( \hspace{-.2cm}
\begin{array}{c}
s_{1\lambda } \\
s_{2\lambda }
\end{array}
\hspace{-.2cm}\right) =\left( \hspace{-.2cm}
\begin{array}{ll}
\lambda /4-z_1\left( t\right) /2 & 
\lambda /4+\left( \mu
+Kd\left( t\right) \right) /2 \\
\lambda /4+\left( \mu +Kd\left( t\right) \right) /2 & 
\lambda /4- 
z_2\left( t\right) /2
\end{array}
\hspace{-.2cm}\right) \left( \hspace{-.2cm}
\begin{array}{c}
s_{1\lambda } \\
s_{2\lambda }
\end{array}
\hspace{-.2cm}\right) +\left( \hspace{-.2cm}
\begin{array}{c}
h_{1\lambda }\left( t\right) +\xi _{1\lambda }\left( t\right)  \\
h_{2\lambda }\left( t\right) +\xi _{2\lambda }\left( t\right)
\end{array}
\hspace{-.2cm}\right)
\label{Laneq}
\end{equation}
where $z_a\left( t\right) $ ($a=1,2$) are two Lagrange
multipliers enforcing the spherical constraints, $h_{a\lambda }\left(
t\right) $ two external fields
and $\xi _{a\lambda }\left( t\right) $ the thermal noises with zero mean and correlations $%
\left\langle \xi _{a\lambda }\left( t\right) \xi _{b\mu }\left( t^{\prime
}\right) \right\rangle =2T\delta _{ab}\delta _{\lambda \mu }\delta \left(
t-t^{\prime }\right) $. The initial fields $s_{a\lambda
}\left( 0\right) $ are assumed to be Gaussian variables
with zero mean and variance $\overline{s_{a\lambda }\left(
0\right) s_{b\lambda }\left( 0\right) }=1+2\left( 1-\delta _{ab}\right)
\left( d\left( 0\right) -1\right) $ where $d\left( 0\right) $ is the initial
density. We study the correlation functions $%
C_{1a}\left( t,t^{\prime }\right) =$ $\int d\lambda \,\rho \left( \lambda
\right) \left\langle s_{1\lambda }\left( t\right) s_{a\lambda }\left(
t^{\prime }\right) \right\rangle $ ($a=1,2$), the response functions $%
G_{1a}\left( t,t^{\prime }\right) =\int d\lambda \,\rho \left( \lambda
\right) \delta \left\langle s_{1\lambda }\left( t\right) \right\rangle 
/\delta h_{a\lambda }\left( t^{\prime }\right) $ and the density $d\left(
t\right) =\left( C_{12}\left( t,t\right) +1\right) /2$. Due to its
linearity, the Langevin system (\ref{Laneq}) can be explicitly solved for
given noises, thus the dynamical quantities can be evaluated averaging over
the noises and the eigenvalues $\lambda $ \cite{CCN,CdP}. We find for $%
t\geq t^{\prime }$ and zero external fields:
\begin{equation}
\begin{array}{l}
C_{1a}\left( t,t^{\prime }\right) =\dfrac{\Gamma \left( \frac{t+t^{\prime }}2%
\right) }{\sqrt{\Gamma \left( t\right) \Gamma \left( t^{\prime }\right) }}%
\left[ d\left( \dfrac{t+t^{\prime }}2\right) -2T
\dint_{t^{\prime }}^{\left(
t+t^{\prime }\right) /2}du\dfrac{I_1\left( t+t^{\prime }-2u\right) }{%
t+t^{\prime }-2u}\dfrac{\Gamma \left( u\right) }{\Gamma \left( \frac{%
t+t^{\prime }}2\right) }\right]  \\ 
\quad \quad \quad \quad \quad +\eta _a\dfrac{\Sigma \left( \frac{t+t^{\prime }%
}2\right) }{\sqrt{\Sigma \left( t\right) \Sigma \left( t^{\prime }\right) }}%
\left[ 1-d\left( \dfrac{t+t^{\prime }}2\right) -T\dint_{t^{\prime }}^{\left(
t+t^{\prime }\right) /2}du\,\dfrac{\Sigma \left( u\right) }{\Sigma \left( 
\frac{t+t^{\prime }}2\right) }\right]  \\ 
G_{1a}\left( t,t^{\prime }\right) =\sqrt{\dfrac{\Gamma \left( t^{\prime
}\right) }{\Gamma \left( t\right) }}\dfrac{I_1\left( t-t^{\prime }\right) }{%
t-t^{\prime }}+\dfrac{\eta _a}2\sqrt{\dfrac{\Sigma \left( t^{\prime }\right) }{%
\Sigma \left( t\right) }}
\end{array}
\label{twotimequantities}
\end{equation}
where $\eta _a=\left( -1\right) ^{a+1}$, $I_1\left( t\right) $ is the
modified Bessel function of order $1$ and $\Sigma \left( t\right) =\Gamma
\left( t\right) e^{2K\int_0^td\left( u\right) du+2\mu t}$, while $\Gamma
\left( t\right) $ and $d\left( t\right) $ satisfy the following coupled
integro-differential equations: 
\begin{equation}
\begin{array}{l}
\Gamma \left( t\right) d\left( t\right) =d\left( 0\right) \dfrac{I_1\left(
2t\right) }t+T\dint_0^tdu\dfrac{I_1\left( 2\left( t-u\right) \right) }{t-u}%
\Gamma \left( u\right) \\ 
\dfrac d{dt}d\left( t\right) =\left( 1-d\left( t\right) \right) \left( \dfrac
d{dt}\ln \Gamma \left( t\right) +2\mu +2Kd\left( t\right) \right) -T
\label{integsyst}
\end{array}
\end{equation}
with the initial conditions $\Gamma \left( 0\right) =1$ and arbitrary $%
d\left( 0\right) $.

The special case $K=0$ is analytically tractable by
standard Laplace techniques \cite{CCN} and one can find the full time
dependence of $\Gamma $ and $d$, but this is not the case for $K>0$.
However, it can be
verified that the long time fixed points of the $d\left( t\right) $-equation 
are described by the $P$ and $SG$ equilibrium equations (\ref{P-sd.eq2})
with $\Gamma \left( t\right) $ respectively proportional to $%
\,e^{\left( z/\beta \right) t}$ and $e^{2t}/\sqrt{4\pi t^3}$;
furthermore local analysis of these
fixed points yields the same stability conditions of the equilibrium theory.
Moving from high temperature towards the equilibrium second order line, 
a diverging equilibration time $\tau _{eq}^{2nd}\sim \left( T-T_c\right) ^{-2}$ is found \cite{CCN}. 
In the (both $P/SG$ and $P/P$) first order region, between the spinodal
lines in the plane $\mu -T$, two distinct asymptotic solutions are possible;
to know which of them is established for long times,
stability analysis only is not sufficient, but an appropriate
matching with the initial conditions is required.
This has been achieved by
numerically solving Eqs. (\ref{integsyst}).
As an example,
the different curves in Fig. 2 are obtained by keeping $d\left( 0\right) ,K,T$ fixed and
taking several values of $\mu $ slightly above and below the
estimated critical threshold $\mu _d$ at which the system jumps from the $P$ to the  
$SG$ solution. The approach to the critical threshold of
the density function is accompanied by a diverging plateau
($\tau _{eq}^{1st}\sim -\ln \left( \mu _d-\mu \right) $)
at an unstable fixed point given by the further root at the $SG$
of Eqs. (\ref{P-sd.eq2}), before falling in one of
the two stable asymptotic values. Fig. 3 shows that at fixed $K$ the whole
critical line depends on the value of the
initial density $d\left( 0\right) $.
Lower values of $d\left( 0\right) $
favor the $P$ phase and, for very low values, the
$SG$ solution is obtained only when the $P$ solution becomes
unstable, i.e. at the $P$ spinodal line; in this last case
$\tau _{eq}^{1st}\sim \left( \mu _{P-sp}-\mu \right) ^{-1}$.
A such behavior is to be essentially related to the competition of the two
phases in the first order region; the same features are found for the
$P/P$ discontinuous transition, apart from the power law behavior of 
$\Gamma \left( t\right) e^{-2t}$ visible in Fig. 2.   

The equilibrium dynamics in the $P$ phase is formally equivalent 
to a trivial schematic formulation of the Mode Coupling Theory (MCT) with two modes $a=1,2$
(using the notation of \cite{Gotze} the MCT kernel is $F_1$); the two-time quantities 
(\ref{twotimequantities})
are functions of $t-t^{\prime }$
and satisfy the fluctuation-dissipation theorem (FDT), 
decaying 
to zero with relaxation time
$\tau _{eq}^{2nd}$ \cite{CCN}. 
In the $SG$ phase, due to the power law factor in 
$\Gamma \left( t\right) $, a
nonequilibrium long-time dynamics is found; this is characterized
by a regime ($t\simeq t^{\prime }$) where
FDT holds, followed by an aging regime ($t>t^{\prime }$) where
the correlation $C_{1a}$ is a function of $t^{\prime }/t$ and 
the integrated response
$\chi _{1a}\left(t,t^{\prime }\right) =\int _{t^{\prime }}^tG_{1a}
\left( t,u\right) du$
plotted as a function of the correlation, $\chi _{1a}\left( C_{1a}\right )$,
is constant
\cite{CCN,CdP}. 
We notice that a $SG$-like nonequilibrium regime is present
also in the $P$ phase, close enough to the $P/SG$ 
discontinuous transition line, 
for waiting times lower than $\tau _{eq}^{1st}$ (Fig. 2).
In this case one can well distinguish the following three steps in the relaxation
process (see the curves $T\chi _{11}\left( C_{11}\right) $ in Fig. 4):
an initial FDT regime ($\tau _{eq}^{1st}>t\simeq t^{\prime }$%
) followed by an aging regime ($\tau _{eq}^{1st}>t>t^{\prime }$), as for the
$SG$ phase; finally an interrupted aging regime ($t>\tau
_{eq}^{1st}>t^{\prime }$).

In conclusion, we have analyzed the equilibrium and dynamical properties of
the spherical frustrated BEG model at mean field level for $K>0$. 
The introduction of dilution and attraction
in a standard spin glass model generates a 
coexistence line between the $P$ and $SG$ phase that exhibits
new interesting features.
Future extensions or improvements of our framework can be realized,
such as a version of the model with $p>2$-particles interactions \cite{CCN1};
this last case 
could display interesting connections with recent studies 
on the MCT for glassy systems with attractive interactions
\cite{MCT}. 
\vspace{0.9cm}

\begin{figure}[h!]
\begin{center}
\vspace{-1.1cm}
\epsfxsize=5cm\epsfbox{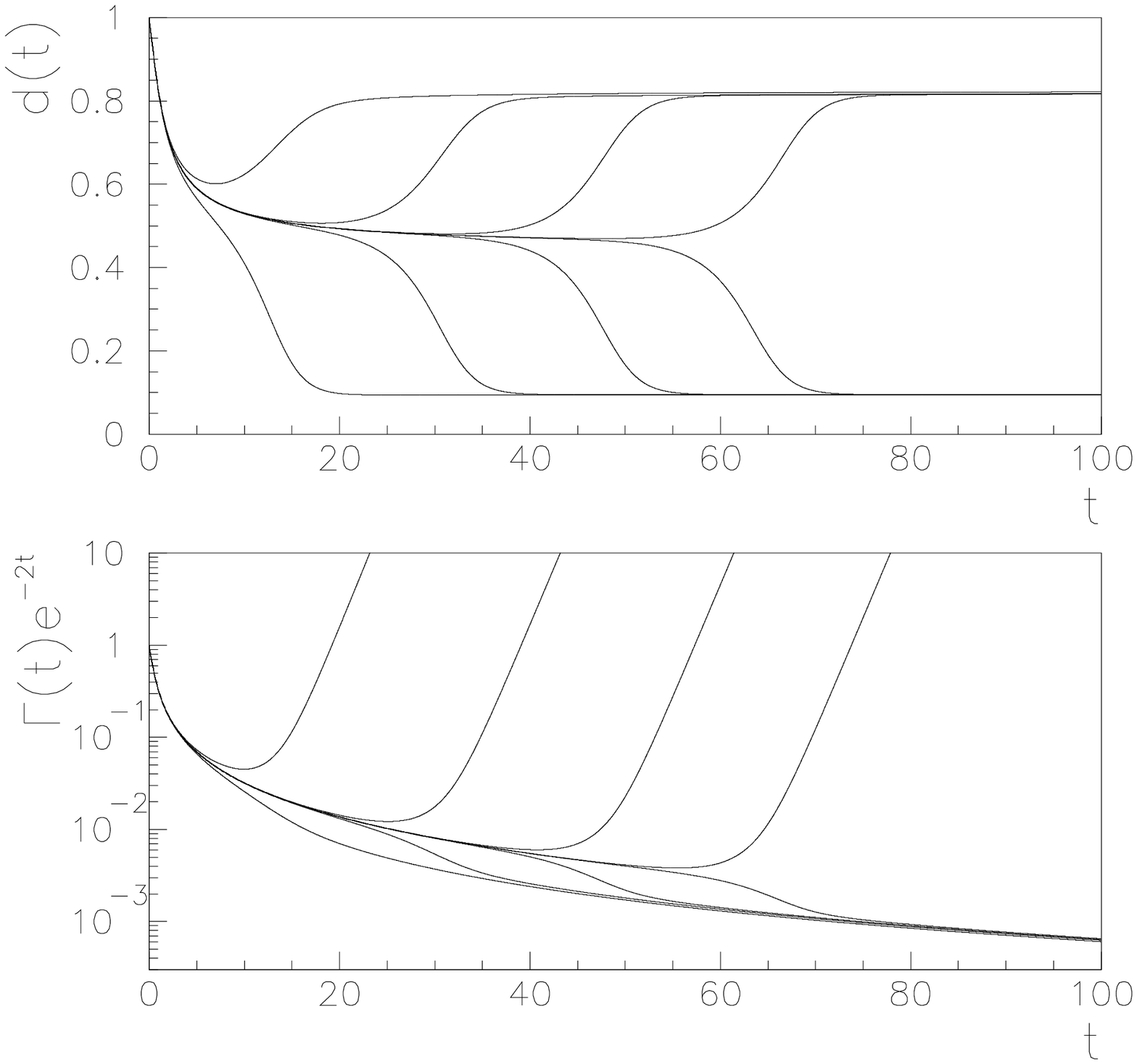}
\hspace{-0.55cm}
\epsfxsize=5cm\epsfbox{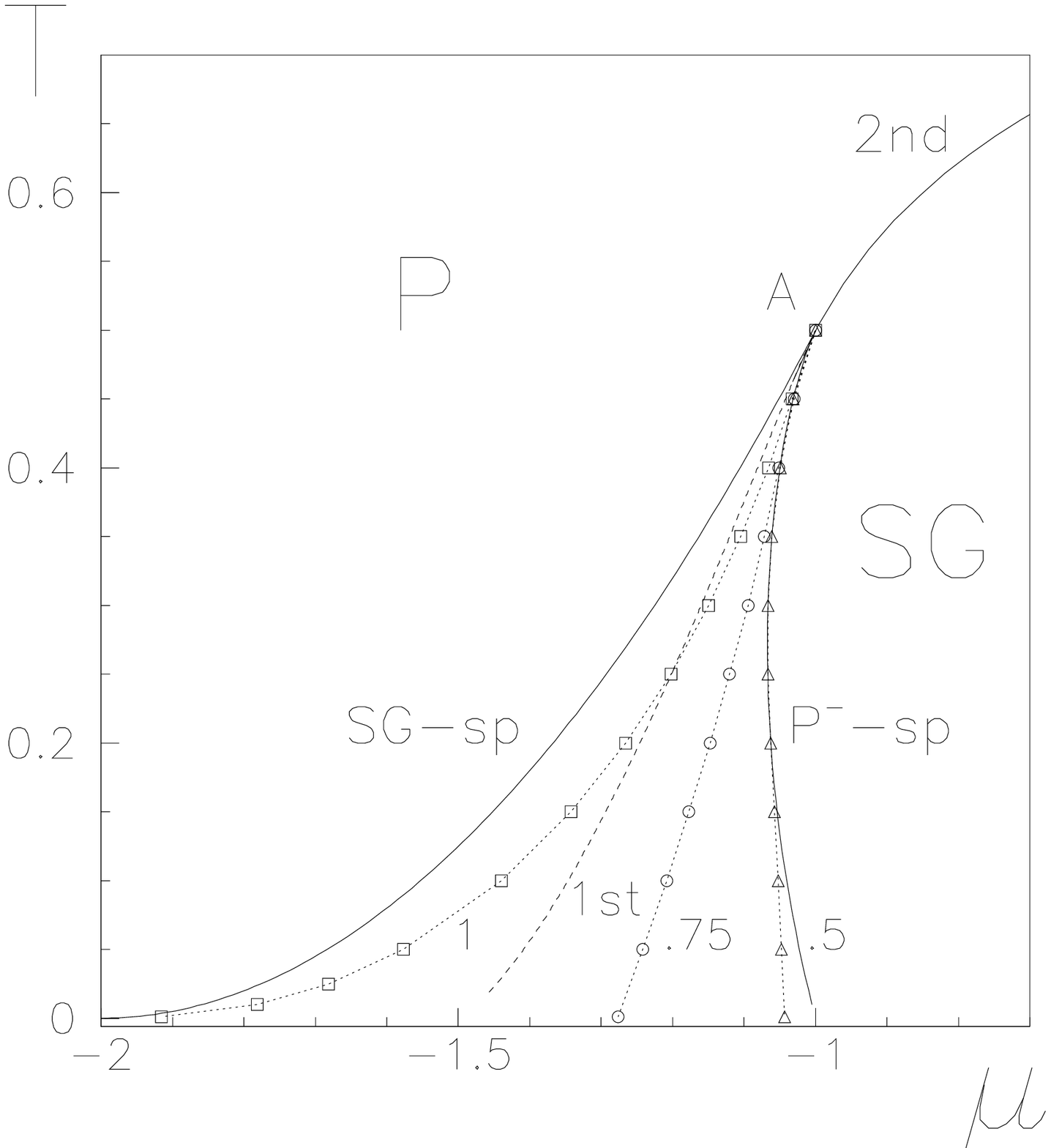}
\hspace{-0.6cm}
\epsfxsize=5cm\epsfbox{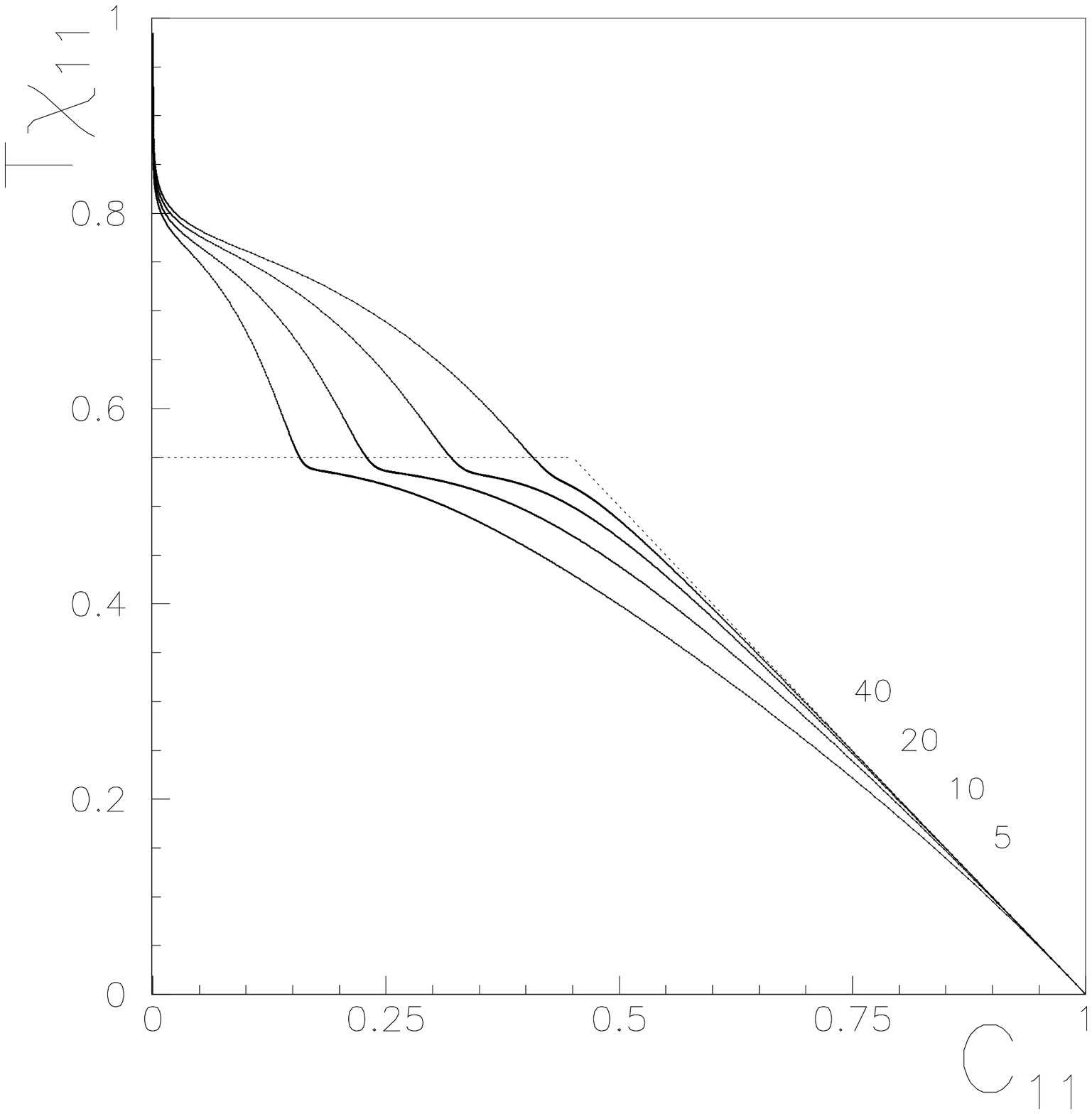}
\vspace{-1.1cm}
\end{center}
\hspace{2.2cm}Fig. 2\hspace{3.7cm}Fig. 3\hspace{3.75cm}Fig. 4
\vspace{-.2cm}
\caption{\footnotesize
Graphs of density $d\left( t\right) $ and $\Gamma \left( t\right) e^{-2t}$
(in logarithmic scale) when the system is close to the discontinuous $P/SG$ transition.
The curves are obtained by numerically solving Eqs. (\ref
{integsyst}) for $\mu -\mu _d=\pm 10^{-x}$ with $x=2,4,6,8$ 
from left to right and $\mu
_d\simeq -1.27$; the other parameters are
fixed at $K=1$, $T=0.2$ and $d\left( 0\right) =1$.
Before falling in one of the two stable asymptotic values the density function
exhibits
a plateau at an unstable intermediate fixed point,
whose length diverges as $\mu $ approaches 
the critical threshold $\mu _d$ ($\tau _{eq}^{1st} \sim -\ln \left(\mu _d-\mu \right) $).
}
\vspace{-0.2cm}
\caption{\footnotesize 
We show the $P/SG$ discontinuous critical lines for three values of the initial density,  
$d\left( 0\right) =1,$ $0.75,$ $0.5$ (squares, circles, triangles),
in the region of the plane $\mu -T$ between the spinodal
lines ($K=1$). The equilibrium first order 
line is also shown (dashed line). 
For $d\left( 0\right) $ lower than $\simeq 0.4$ the critical
line coincides with the $P$ spinodal line.}
\vspace{-0.2cm}
\caption{\footnotesize
We plot $T\chi _{11}$ vs. $C_{11}$ for several values of $t^{\prime }$
(indicated in the figure) when
the system is
in the $P$ phase but very close to the $P/SG$ discontinuous transition line 
($K=1$, $T=0.1$,
$d\left( 0\right) =1,\mu \simeq
-1.44$)
and $t^{\prime }<\tau _{eq}^{1st}\simeq 100$. One
can distinguish in the relaxation process
an initial FDT regime ($T\chi _{11}=1-C_{11}$),
followed by an aging regime, as for the $SG$ phase,
where the curves approach 
the dotted line
($T\chi _{11}=1+T-d$ where $d$ is the density at
the unstable fixed point);
finally an interrupted aging regime,
due to the finite equilibration time,
in which the curves bend upward. }
\label{integ-d-g}
\end{figure}
\vspace{-.2cm}
\acknowledgments
This work has been partially supported by MIUR-PRIN 2002 and MIUR-FIRB 2002.


\begin{thebibliography}{99}
\bibitem{replica}  This is equivalent to a replica symmetric solution in the
replica framework;
the replica free energy has to be maximized
with respect to the overlap $q$ and minimized with respect to the density $d$
\cite{KTJ,CCN1}.
\end{thebibliography}
\end{document}